\PassOptionsToPackage{unicode}{hyperref}
\PassOptionsToPackage{hyphens}{url}
\documentclass[
]{article}
\usepackage{xcolor}
\usepackage{amsmath,amssymb}
\usepackage{iftex}
\ifPDFTeX
  \usepackage[T1]{fontenc}
  \usepackage[utf8]{inputenc}
  \usepackage{textcomp} 
\else 
  \usepackage{unicode-math} 
  \defaultfontfeatures{Scale=MatchLowercase}
  \defaultfontfeatures[\rmfamily]{Ligatures=TeX,Scale=1}
\fi
\usepackage{lmodern}
\ifPDFTeX\else
\fi
\IfFileExists{upquote.sty}{\usepackage{upquote}}{}
\IfFileExists{microtype.sty}{
  \usepackage[]{microtype}
  \UseMicrotypeSet[protrusion]{basicmath} 
}{}
\makeatletter
\@ifundefined{KOMAClassName}{
  \IfFileExists{parskip.sty}{%
    \usepackage{parskip}
  }{
    \setlength{\parindent}{0pt}
    \setlength{\parskip}{6pt plus 2pt minus 1pt}}
}{
  \KOMAoptions{parskip=half}}
\makeatother
\usepackage{color}
\usepackage{fancyvrb}

\DefineVerbatimEnvironment{Highlighting}{Verbatim}{commandchars=\\\{\}}
\newenvironment{Shaded}{}{}

\newcommand{\AttributeTok}[1]{\textcolor[rgb]{0.49,0.56,0.16}{#1}}

\newcommand{\BuiltInTok}[1]{\textcolor[rgb]{0.00,0.50,0.00}{#1}}

\newcommand{\ControlFlowTok}[1]{\textcolor[rgb]{0.00,0.44,0.13}{\textbf{#1}}}
\newcommand{\DataTypeTok}[1]{\textcolor[rgb]{0.56,0.13,0.00}{#1}}
\newcommand{\DecValTok}[1]{\textcolor[rgb]{0.25,0.63,0.44}{#1}}

\newcommand{\FunctionTok}[1]{\textcolor[rgb]{0.02,0.16,0.49}{#1}}

\newcommand{\KeywordTok}[1]{\textcolor[rgb]{0.00,0.44,0.13}{\textbf{#1}}}
\newcommand{\NormalTok}[1]{#1}
\newcommand{\OperatorTok}[1]{\textcolor[rgb]{0.40,0.40,0.40}{#1}}

\newcommand{\StringTok}[1]{\textcolor[rgb]{0.25,0.44,0.63}{#1}}

\usepackage{longtable,booktabs,array}
\usepackage{calc} 
\usepackage{etoolbox}
\makeatletter
\patchcmd\longtable{\par}{\if@noskipsec\mbox{}\fi\par}{}{}
\makeatother
\IfFileExists{footnotehyper.sty}{\usepackage{footnotehyper}}{\usepackage{footnote}}
\makesavenoteenv{longtable}
\providecommand{\tightlist}{%
  \setlength{\itemsep}{0pt}\setlength{\parskip}{0pt}}
\usepackage[]{natbib}
\usepackage{bookmark}
\IfFileExists{xurl.sty}{\usepackage{xurl}}{} 
\hypersetup{
  hidelinks,
  pdfcreator={LaTeX via pandoc}}

\author{}
\date{}

\begin{document}

\section{ReqToCode: Embedding Requirements Traceability as a Structural
Property of the
Codebase}\label{reqtocode-embedding-requirements-traceability-as-a-structural-property-of-the-codebase}

\textbf{Thorsten Schlathölter, Dr.}

\emph{Independent Researcher}

\emph{thorsten@ariadne-thread.io}

\begin{center}\rule{0.5\linewidth}{0.5pt}\end{center}

\subsection{Abstract}\label{abstract}

Requirements traceability in safety-critical software development
remains largely dependent on external documentation maintained
separately from the systems it describes. This separation introduces
structural fragility: traces degrade silently as requirements, code, and
tests evolve independently across tools, repositories, and revisions.
Recent advances in LLM-based traceability focus on recovering broken
links after the fact --- an inherently retrospective approach. This
paper introduces \textbf{ReqToCode}, an approach that prevents trace
degradation by embedding traceable system elements directly into the
codebase, making traceability a compile-time verifiable property of the
system rather than an external documentation task. Central to the
approach is the concept of the \textbf{Traceable} --- a language-native,
generated code element that represents a single requirement and carries
its metadata. Developers reference Traceables in implementation and test
code, creating hard, bidirectional links that are validated
automatically during the build process. When requirements change, the
system responds through a graduated lifecycle --- from deprecation
warnings to build failures --- providing teams with actionable signals
rather than abrupt breakage. We describe the approach, its architectural
principles, the Traceable lifecycle, and illustrate it with a generic
example spanning requirement definition, artifact generation, code
integration, and build-time validation.

\textbf{Keywords:} requirements traceability, safety-critical software,
compile-time verification, code generation, bidirectional traceability,
regulated industries

\begin{center}\rule{0.5\linewidth}{0.5pt}\end{center}

\subsection{1. Introduction}\label{introduction}

In regulated software development --- automotive (ISO 26262), medical
devices (IEC 62304), aerospace (DO-178C), and similar domains ---
requirements traceability is not optional. Standards mandate that every
requirement be traceable to its implementation and verification, and
that this traceability be demonstrable during audits and assessments.

In practice, traceability is typically maintained in external artifacts:
spreadsheets, traceability matrices, or dedicated fields in Application
Lifecycle Management (ALM) tools such as Jira, Codebeamer, IBM DOORS, or
Siemens Polarion. These artifacts exist outside the codebase and are
maintained manually or semi-manually by development teams.

This separation creates a fundamental problem. As systems evolve, the
external traceability record and the actual system diverge. Requirements
are added, renamed, or deleted in the ALM tool. Code is refactored,
moved, or rewritten in the repository. Tests change scope or are
restructured. At no point does the system itself enforce consistency
between these artifacts and the traceability record. The result is a
traceability debt that accumulates silently and surfaces --- often
painfully --- during audits, assessments, or safety reviews. A recent
empirical study confirms this picture: 80\% of surveyed practitioners
identified cost as the primary barrier to traceability adoption, with
manual maintenance burden cited as a dominant contributor {[}1{]}.

The emergence of AI-assisted development intensifies this problem. As
code generation tools produce increasing volumes of implementation and
test code, the gap between what exists in the codebase and what is
documented in traceability records widens further. Generated code must
be explainable, reviewable, and traceable to the same standard as
handwritten code, yet current practices offer no structural mechanism to
ensure this.

Recent research has responded to the traceability challenge primarily
through recovery-based approaches, leveraging large language models
(LLMs) to reconstruct trace links after they have degraded or gone
missing {[}2, 3, 4{]}. While these approaches achieve impressive
accuracy, they are fundamentally retrospective: they detect or
reconstruct traces rather than preventing their degradation. A
complementary line of work has begun exploring how to embed traceability
into LLM code generation pipelines themselves {[}5{]}, recognizing that
trace links must be established at creation time rather than recovered
later.

This paper proposes \textbf{ReqToCode}, an approach that addresses the
traceability problem at its structural root. Rather than maintaining
traces externally or recovering them after the fact, ReqToCode generates
language-native artifacts --- termed \textbf{Traceables} --- directly
from authoritative requirement sources. Developers reference these
Traceables in their implementation and test code, creating hard links
that are verified at compile time. When requirements change, a
deprecation-to-removal lifecycle provides teams with warnings before
build failures --- rather than abrupt breakage.

The remainder of this paper is structured as follows. Section 2 examines
related work and the limitations of existing traceability approaches.
Section 3 describes the ReqToCode approach in detail, including the
Traceable concept and its lifecycle. Section 4 presents a generic
example illustrating the complete workflow. Section 5 discusses
properties, limitations, and implications. Section 6 concludes.

\begin{center}\rule{0.5\linewidth}{0.5pt}\end{center}

\subsection{2. Related Work}\label{related-work}

\subsubsection{2.1 Traditional Traceability
Approaches}\label{traditional-traceability-approaches}

Requirements traceability has been studied extensively since Gotel and
Finkelstein's foundational work on traceability problem characterization
{[}6{]}. Cleland-Huang et al.~{[}10{]} provide a comprehensive overview
of the field's evolution, identifying key research areas including
automated trace creation, trace maintenance, and the pursuit of
ubiquitous traceability.

The dominant paradigm relies on traceability matrices --- tabular
mappings between requirements and downstream artifacts --- maintained
either manually or within ALM tools. Tools such as IBM DOORS, Siemens
Polarion, PTC Codebeamer, and Jama Connect provide built-in traceability
fields that allow users to create links between requirements and other
artifacts. These links, however, are metadata within the ALM system.
They describe intended relationships but do not participate in the
development workflow in a way that enforces consistency. A developer may
delete a function that implements a requirement without any signal from
the traceability system.

Ruiz et al.~{[}1{]} provide empirical evidence for the practical
consequences of this separation. Their study of 55 practitioners found
that traceability is still widely perceived as a costly, manual
activity, with rapid iteration cycles in agile environments frequently
breaking manual traces and rendering traditional matrices obsolete.

\subsubsection{2.2 Traceability Maintenance and
Decay}\label{traceability-maintenance-and-decay}

Mäder and Gotel {[}11{]} address the problem of traceability decay ---
the progressive degradation of trace links as systems evolve --- through
semi-automated maintenance strategies that recognize model changes and
apply rule-based updates to existing trace relations. Their work
demonstrates that traceability maintenance can be partially automated
but remains dependent on continuous monitoring and human verification.

Mäder and Egyed {[}12{]} provide empirical evidence that developers with
access to maintained trace links perform maintenance tasks significantly
faster and with higher correctness. This result underscores that the
value of traceability is contingent on its currency --- stale links are
not merely useless but actively harmful when they mislead developers
about the system's actual structure.

These findings directly motivate the ReqToCode approach: if traceability
degrades because trace links are metadata that can silently fall out of
sync with the code, then preventing degradation requires making traces
part of the code itself. Unlike event-based traceability {[}10{]}, which
monitors changes in artifacts and updates trace metadata accordingly,
ReqToCode eliminates metadata links entirely by embedding requirement
identities directly into the type system.

\subsubsection{2.3 Information Retrieval and Deep Learning for
Traceability}\label{information-retrieval-and-deep-learning-for-traceability}

Automated traceability link recovery has progressed through several
generations of techniques. Early approaches applied information
retrieval methods such as vector space models and latent semantic
indexing to match textual similarity between requirements and code
artifacts {[}8{]}.

Guo et al.~{[}13{]} advanced the field by applying deep learning to
traceability, using recurrent neural networks to capture semantic
relationships between requirements and design documents that textual
similarity alone cannot detect. This work opened a productive line of
research that continued through BERT-based and transformer-based
approaches.

Most recently, large language models have been applied to traceability
link recovery (TLR). Niu et al.~{[}2{]} introduce TVR, a
retrieval-augmented generation (RAG) approach for validating and
recovering trace links between stakeholder and system requirements in
automotive systems, achieving 98.87\% validation accuracy on industrial
data. Hey et al.~{[}3{]} propose LiSSA, a generic RAG-based approach for
inter-requirements traceability link recovery, evaluated across multiple
benchmark datasets. Rodriguez et al.~{[}7{]} explore prompt engineering
strategies for extracting trace link predictions from LLMs,
demonstrating that prompt design significantly affects recovery quality.
Hassine {[}4{]} applies LLMs specifically to recovering trace links
between security requirements and goal models, illustrating the breadth
of domains where recovery approaches are being explored.

Despite significant improvements in accuracy, all recovery-based
approaches --- from LSI through deep learning to LLMs --- share a
fundamental characteristic: they operate retrospectively. They recover
or validate traces that have already degraded, rather than preventing
degradation structurally. Their output is probabilistic --- candidate
links that require human validation --- and they cannot guarantee
completeness.

\subsubsection{2.4 Embedding Traceability in AI-Generated
Code}\label{embedding-traceability-in-ai-generated-code}

Wang et al.~{[}5{]} recognize that the rise of LLM-based code generation
creates a new category of traceability challenge and propose embedding
traceability as a first-class objective in LLM code generation
pipelines. Their framework addresses LLM opacity and non-determinism
through structured requirement prompting, metadata-aware fine-tuning,
and retrieval-augmented validation.

This work shares ReqToCode's premise that traceability should be
addressed at creation time rather than recovered later. However, it
focuses on making LLMs produce traceable code, whereas ReqToCode
provides the structural foundation --- the Traceables --- that both
human developers and AI tools can reference. The two approaches are
complementary: an LLM generating code within a ReqToCode-enabled project
would reference the same Traceables as a human developer, with the same
compile-time guarantees.

\subsubsection{2.5 Annotation-Based
Approaches}\label{annotation-based-approaches}

Some approaches embed traceability information directly in source code
through annotations or comments. Antoniol et al.~{[}14{]} established
the connection between code documentation and requirements through
information retrieval techniques applied to comments and identifiers,
and subsequent work has explored comment-based and annotation-based
trace embedding as a lightweight alternative to external matrices.

These approaches move traceability closer to the code but suffer from a
critical limitation: the annotations are unverified strings. If a
requirement identifier is misspelled, renamed, or deleted, the
annotation remains in the code without producing any error. There is no
structural link between the annotation and an authoritative source of
requirement definitions.

\subsubsection{2.6 Model-Driven
Approaches}\label{model-driven-approaches}

Model-driven engineering (MDE) approaches establish traceability through
formal models and transformations {[}9{]}. While these provide
structural rigor, they require adoption of model-driven workflows that
are often incompatible with the code-centric practices of modern
software development teams. Teams working with Git-based workflows,
continuous integration, and iterative development rarely adopt full MDE
tool chains.

\subsubsection{2.7 Gap}\label{gap}

The current research landscape addresses traceability primarily through
recovery (Section 2.3), maintenance of existing links (Section 2.2), or
embedding trace-awareness into AI generation pipelines (Section 2.4).
Recovery and maintenance approaches operate on trace links represented
as metadata relations and invest in repairing them after they degrade.
The gap addressed by ReqToCode is the absence of an approach that:

\begin{enumerate}
\def\labelenumi{\arabic{enumi}.}
\tightlist
\item
  Creates \textbf{hard, verifiable links} between requirements and code
  --- not metadata, not annotations, not probabilistic matches.
\item
  Operates within \textbf{existing development workflows} --- Git,
  CI/CD, standard build tools --- without requiring adoption of
  specialized tool chains.
\item
  Responds to requirement changes through a \textbf{graduated lifecycle}
  --- providing proportional signals from deprecation warnings to build
  failures, rather than binary breakage.
\item
  Supports \textbf{bidirectional change detection} --- identifying when
  requirements change without corresponding code updates, and when code
  changes without corresponding requirement updates.
\item
  Enables \textbf{branch-scoped traceability} --- allowing coverage
  analysis to distinguish between baseline and branch-specific
  contributions across parallel development contexts.
\item
  Provides \textbf{coverage analysis at any revision} --- enabling teams
  to determine which requirements are referenced in implementation and
  tests at any point in the version control history.
\item
  Provides a \textbf{structural foundation} that both human developers
  and AI code generation tools can reference equally.
\end{enumerate}

\begin{center}\rule{0.5\linewidth}{0.5pt}\end{center}

\subsection{3. The ReqToCode Approach}\label{the-reqtocode-approach}

\subsubsection{3.1 Core Idea}\label{core-idea}

ReqToCode inverts the traditional traceability model. Rather than
maintaining traces as external metadata that describes the system,
ReqToCode generates \textbf{language-native artifacts} from
authoritative requirement sources and embeds them directly in the
codebase. These artifacts become part of the system's type system,
compilation unit, or equivalent language-level construct. Developers
reference them in implementation and test code just as they would
reference any other code element.

The consequence is that traceability links become subject to the same
verification mechanisms as any other code dependency: the compiler, the
linker, or the build system. A broken trace is a broken build.

\subsubsection{3.2 The Traceable}\label{the-traceable}

Central to the ReqToCode approach is the concept of the
\textbf{Traceable}: a generated, language-native code element that
represents a single requirement and serves as the anchor point for all
trace links to that requirement.

A Traceable is:

\begin{itemize}
\tightlist
\item
  \textbf{Generated}, not manually authored. It is derived from an
  authoritative requirement source and regenerated on every
  synchronization cycle. Manual modifications are overwritten.
\item
  \textbf{Language-native}. It is expressed in the target language's
  type system --- as an enumeration constant, a typed definition, a
  structured element, or an equivalent construct that the compiler can
  verify.
\item
  \textbf{Metadata-carrying}. It embeds the requirement's title, status,
  version, and other properties as structured data accessible at both
  compile time and runtime.
\item
  \textbf{Referenceable}. Developers import and reference Traceables in
  implementation and test code using standard language constructs:
  imports, type references, method calls.
\item
  \textbf{Lifecycle-aware}. A Traceable reflects the lifecycle state of
  its source requirement, including intermediate states such as
  deprecation (see Section 3.5).
\end{itemize}

The term ``Traceable'' is used deliberately to distinguish these
artifacts from the requirements they represent. A requirement exists in
the ALM tool; a Traceable exists in the codebase. The Traceable is a
projection of the requirement into the developer's domain --- a
compile-time verifiable anchor that makes the requirement referenceable
in code.

\subsubsection{3.3 RequirementSets}\label{requirementsets}

In real-world systems, the number of requirements can reach into the
hundreds or thousands. Generating all Traceables into a single module
would produce an unwieldy artifact that is difficult to review and
conflates unrelated concerns.

ReqToCode addresses this through \textbf{RequirementSets} --- named,
scoped groupings of Traceables that partition the requirement space into
coherent units. Each RequirementSet corresponds to a defined subset of
requirements from the source system and generates a separate,
self-contained code module.

The grouping criteria are defined by the project and typically reflect
the structure already present in the ALM tool or the system
architecture. Common partitioning strategies include:

\begin{itemize}
\tightlist
\item
  \textbf{By requirement type.} Stakeholder requirements, system
  requirements, software requirements, and architecture requirements
  each form their own RequirementSet. This mirrors the hierarchical
  decomposition mandated by standards such as ASPICE and ISO 26262.
\item
  \textbf{By component or subsystem.} A sensor validation module
  references only the RequirementSet for sensor-related requirements,
  not the full system scope. This keeps imports focused and compilation
  units small.
\item
  \textbf{By source.} When a project integrates requirements from
  multiple ALM tools or structured repositories, each source can map to
  its own RequirementSet.
\end{itemize}

Each Traceable belongs to exactly one RequirementSet. If a requirement
is relevant to multiple components, the partitioning strategy should
place it at the appropriate architectural level --- typically a
system-level RequirementSet --- rather than duplicating it across
component-level sets.

RequirementSets serve two purposes. First, they reduce cognitive
overhead by ensuring that developers only import the Traceables relevant
to their component. Second, they provide a natural unit for coverage
analysis (Section 3.7): coverage can be reported per RequirementSet,
making it immediately visible which areas of the system have strong
traceability and which have gaps.

Each RequirementSet is generated as a separate, self-contained code
module and committed to the repository alongside the project's source
code.

\subsubsection{3.4 Architectural Overview}\label{architectural-overview}

The ReqToCode approach operates through a pipeline with the following
stages:

\textbf{Source synchronization.} An authoritative requirement source ---
an ALM tool, a structured file repository, or any system that manages
requirement definitions --- is periodically read to obtain the current
set of requirements and their metadata.

\textbf{Traceable generation.} From the synchronized requirements, the
system generates source code artifacts in the target language, organized
into RequirementSets (Section 3.3). Each requirement is represented as a
Traceable --- a named, typed element carrying its metadata. Each
RequirementSet forms a self-contained module that can be imported
independently by code in the project.

\textbf{Codebase integration.} The generated Traceables are committed to
the project's version control system through the standard development
workflow. This typically involves creating a branch, writing the
artifacts, and submitting the changes for review through the platform's
merge or pull request mechanism. Developers review and accept the
generated changes as they would any other code contribution.

\textbf{Reference and verification.} Developers reference Traceables in
their implementation and test code. These references are standard code
dependencies --- imports, type references, constant usages --- that the
language's build tooling verifies automatically. If a referenced
Traceable is removed, all dependent code fails to compile.

\textbf{Change detection.} By comparing the state of requirements
against the state of code references, the system can identify two
critical conditions: (a) a requirement has changed but the referencing
code has not been updated, and (b) code has changed but no corresponding
requirement update exists. Each side uses its natural source of truth:
for requirements, the last-modified timestamp maintained as metadata in
the ALM tool; for code, the commit history in the version control
system. This bidirectional change detection enables continuous
visibility into the alignment between intent and realization. It
requires a mapping between Traceables and the code locations that
reference them --- a mapping that is computable through static analysis
of the codebase, since Traceable references are standard language-level
dependencies resolvable by the compiler or build system.

\subsubsection{3.5 The Traceable
Lifecycle}\label{the-traceable-lifecycle}

In practice, requirements are rarely deleted instantaneously. They move
through lifecycle states in the ALM tool --- from active through various
review and deprecation states to eventual removal. A naive approach that
removes a Traceable the moment its source requirement is deleted would
produce abrupt build failures, disrupting development teams without
warning.

ReqToCode addresses this through a \textbf{graduated lifecycle} for
Traceables that mirrors and extends the requirement lifecycle in the
source system.

\textbf{Active.} The Traceable is generated normally. Code references
compile without warnings. This is the default state for requirements
that are approved and current.

\textbf{Deprecated.} When the source requirement enters a terminal or
removed state --- and the ALM tool provides this lifecycle information
--- the corresponding Traceable is not immediately removed. Instead, it
is marked as deprecated using the target language's deprecation
mechanism (e.g., \texttt{@Deprecated} in Java,
\texttt{{[}{[}deprecated{]}{]}} in C++, or equivalent constructs). The
Traceable remains compilable, but IDEs and build tools surface warnings
at every reference site. This gives development teams visibility into
which code references are affected and time to respond --- by updating,
redirecting, or removing the references.

\textbf{Removed.} After a configurable grace period --- or when the
requirement transitions to a final removal state --- the Traceable is
deleted from the generated artifact. All remaining code references now
produce compilation errors. This is the fail-fast signal that forces
resolution.

The graduated lifecycle serves two purposes. First, it makes ReqToCode
viable in real development environments where requirement changes are
not always coordinated with code changes. Second, it aligns with the
principle that the severity of the signal should be proportional to the
urgency of the response: a recently deprecated requirement warrants
attention, not an emergency.

The availability of deprecation depends on two conditions: the ALM tool
must expose requirement lifecycle transitions, and the target language
must support a deprecation mechanism. Where either condition is not met,
the system falls back to direct removal --- which still provides the
core fail-fast property, albeit without the graduated warning period.

\subsubsection{3.6 Branch-Scoped
Traceability}\label{branch-scoped-traceability}

Because Traceables are code committed to version control, they
participate naturally in the branching model of the underlying version
control system. This enables a property that external traceability
approaches cannot replicate: \textbf{branch-scoped traceability}, where
coverage analysis can distinguish between baseline contributions and
branch-specific work.

In practice, software products are developed in parallel along multiple
axes. A main branch represents the current released or baseline product
with its established set of requirements. Feature branches introduce new
requirements for capabilities under development. Release branches may
carry a subset of requirements targeted for a specific delivery. In
variant-rich product lines --- common in automotive and medical device
development --- different branches may represent product variants with
overlapping but distinct requirement sets.

In a ReqToCode-enabled project, the complete set of Traceables is
generated across all branches, ensuring a single consistent truth
without merge conflicts. Traceables carry metadata --- such as scope
attributes or branch associations --- that enable the system to
distinguish which requirements belong to which development context.

Coverage analysis leverages this metadata to produce
\textbf{branch-relative reports}. On a feature branch, the system can
identify which Traceable references are new compared to the main branch
--- the delta --- and report only the feature's contribution. When
reporting back to the ALM tool, a feature branch reports only the code
references that are not already present on main, providing a clear view
of the feature's traceability state without conflating it with the
baseline.

This has several practical consequences:

\begin{itemize}
\tightlist
\item
  \textbf{No merge conflicts on Traceables.} Because all branches share
  the same complete set, merging a feature branch does not produce
  conflicting Traceable definitions.
\item
  \textbf{Scoped coverage.} Each branch can compute coverage relative to
  its own scope, answering the question: ``How much of \emph{this
  feature's} requirements are traced?'' rather than only ``How much of
  \emph{everything} is traced?''
\item
  \textbf{Incremental ALM reporting.} The ALM tool receives
  branch-specific traceability data, enabling requirement owners to
  track implementation progress per feature or release without noise
  from unrelated branches.
\end{itemize}

External traceability matrices are inherently single-instance --- they
describe one version of the system at a time. Traceables, as committed
code with scope metadata, support parallel development contexts without
duplicating or fragmenting the requirement model.

\subsubsection{3.7 Coverage Analysis}\label{coverage-analysis}

Because Traceables are language-native code elements referenced through
standard imports and method calls, their usage is statically analyzable.
This enables \textbf{requirement coverage analysis at any revision} ---
a capability that traditional traceability approaches can only
approximate through manual effort or periodic audits.

At any commit in the version control history, the system can determine:

\begin{itemize}
\tightlist
\item
  \textbf{Implementation coverage}: which Traceables are referenced in
  production code, and which exist without any implementation reference.
\item
  \textbf{Test coverage}: which Traceables are referenced in test code
  through explicit verification markers (e.g.,
  \texttt{verifiesRequirement(SWR\_101)}), and which lack test
  references.
\item
  \textbf{Lifecycle distribution}: how many Traceables are active, how
  many are deprecated, and how many implementation references target
  deprecated Traceables.
\end{itemize}

This analysis requires no additional tooling beyond what the language
platform already provides. The compiler or build system resolves all
Traceable references; a static analysis pass over the resolved
dependencies produces the coverage data.

The ability to compute coverage at any revision has two practical
consequences. First, it eliminates the audit preparation effort that
teams in regulated industries typically invest weeks in: the
traceability state is computable on demand rather than reconstructed
manually. Second, it enables continuous monitoring --- coverage metrics
can be computed as part of every CI/CD pipeline run, making traceability
degradation visible immediately rather than at the next scheduled
review.

Combined with branch-scoped traceability (Section 3.6), this means
coverage can be computed per branch. A release branch can verify that
all requirements in its scope are covered before delivery. A feature
branch can track how implementation progresses against the feature's
requirements. The main branch provides a continuous baseline of overall
product coverage.

\subsubsection{3.8 Design Principles}\label{design-principles}

The following principles guide the design of a ReqToCode implementation:

\textbf{Compile-time safety.} Traceability violations must manifest as
build failures, not as runtime exceptions, log entries, or report
findings. The build is the primary enforcement mechanism.

\textbf{Non-invasiveness.} The approach must integrate with existing
development workflows, version control practices, and build systems. It
must not require developers to adopt new tools, learn new languages, or
change their day-to-day practices beyond referencing the generated
Traceables.

\textbf{Source authority.} The requirement source (ALM tool or
structured repository) remains the single source of truth. Traceables
are derived, never manually edited. Any manual modification to a
Traceable is overwritten on the next synchronization cycle.

\textbf{Graduated response.} Requirement lifecycle changes should
produce signals proportional to their urgency. Deprecation before
removal gives teams time to respond without disrupting active
development.

\textbf{Fail-fast semantics.} When a Traceable is ultimately removed,
the system must produce an immediate and unambiguous signal. Silent
degradation of traceability is the primary failure mode this approach is
designed to prevent.

\textbf{Complementary positioning.} The approach complements existing
requirement management tools rather than replacing them. It consumes
their data and translates it into a form that participates in the
development and build workflow.

\subsubsection{3.9 Scope and Boundaries}\label{scope-and-boundaries}

ReqToCode addresses the link between requirements and
implementation/test code. It does not replace requirement management,
test management, or configuration management tools. It does not perform
static analysis, code review, or test execution. Its scope is the
structural connection between what is required and what is built --- and
the continuous verification of that connection.

\begin{center}\rule{0.5\linewidth}{0.5pt}\end{center}

\subsection{4. Illustrative Example}\label{illustrative-example}

This section presents a generic example that illustrates the ReqToCode
workflow end to end. The example uses a simplified automotive context
but is intentionally tool-agnostic and language-agnostic in its
description.

\subsubsection{4.1 Requirement Definition}\label{requirement-definition}

Consider a software component responsible for sensor data validation in
a vehicle. Three software requirements are defined in the project's ALM
tool:

{\def\LTcaptype{none} 
\begin{longtable}[]{@{}lll@{}}
\toprule\noalign{}
ID & Title & Status \\
\midrule\noalign{}
\endhead
\bottomrule\noalign{}
\endlastfoot
SWR-101 & Validate sensor range on input & Approved \\
SWR-102 & Reject stale sensor readings & Approved \\
SWR-103 & Log validation failures & Draft \\
\end{longtable}
}

Each requirement has a unique identifier, a title, and a lifecycle
status.

\subsubsection{4.2 Traceable Generation}\label{traceable-generation}

The ReqToCode pipeline reads these requirements and generates a
RequirementSet containing one Traceable per requirement. In this case,
all three requirements belong to the same software requirement category
and are grouped into a single RequirementSet
\texttt{SensorValidation\_SWR}:

\begin{verbatim}
RequirementSet SensorValidation_SWR

  Traceable SWR_101
    title:  "Validate sensor range on input"
    status: APPROVED

  Traceable SWR_102
    title:  "Reject stale sensor readings"
    status: APPROVED

  Traceable SWR_103
    title:  "Log validation failures"
    status: DRAFT
\end{verbatim}

Each Traceable is a named, typed code element within the RequirementSet
module. The generated module is committed to the repository and subject
to standard code review. Developers import only the RequirementSets
relevant to their component.

\subsubsection{4.3 Implementation
Reference}\label{implementation-reference}

A developer implementing the sensor validation logic references the
generated Traceables in their code:

\begin{verbatim}
function validateSensorInput(reading):
    // Implementation of SWR_101
    if reading.value < RANGE_MIN or reading.value > RANGE_MAX:
        trace(SWR_101)
        return INVALID

    // Implementation of SWR_102
    if reading.timestamp < currentTime() - MAX_AGE:
        trace(SWR_102)
        return STALE

    return VALID
\end{verbatim}

The \texttt{trace(SWR\_101)} call is not a comment or an annotation ---
it is a code reference to a generated Traceable. If \texttt{SWR\_101}
does not exist as a generated element, this code does not compile.

\subsubsection{4.4 Test Reference}\label{test-reference}

Similarly, test code references Traceables to establish which
requirements a test verifies:

\begin{verbatim}
test "sensor range validation rejects out-of-range input":
    verifiesRequirement(SWR_101)
    reading = createReading(value: RANGE_MAX + 1)
    result = validateSensorInput(reading)
    assert result == INVALID
\end{verbatim}

The \texttt{verifiesRequirement(SWR\_101)} call creates a compile-time
link between this test and the requirement it validates.

\subsubsection{4.5 Requirement Deprecation --- Graduated
Response}\label{requirement-deprecation-graduated-response}

Suppose \texttt{SWR-102} is marked as deprecated in the ALM tool because
the staleness check is being moved to a different component. On the next
synchronization cycle, the generated Traceable for \texttt{SWR\_102} is
updated:

\begin{verbatim}
Traceable SWR_102 [DEPRECATED]
  title:  "Reject stale sensor readings"
  status: DEPRECATED
\end{verbatim}

The Traceable remains in the codebase. All existing code references
continue to compile. However, IDEs now display deprecation warnings at
every reference site, and build tools may emit warnings depending on
project configuration. The development team has visibility into which
code is affected and can plan the cleanup.

\subsubsection{4.6 Requirement Removal --- Fail
Fast}\label{requirement-removal-fail-fast}

After the grace period expires --- or when \texttt{SWR-102} transitions
to a final removal state in the ALM tool --- the Traceable is deleted
from the generated module.

Every line of code that references \texttt{SWR\_102} --- in
implementation and in tests --- now fails to compile. The development
team is forced to address these references: remove them, redirect them
to a different requirement, or challenge the deletion.

This is the core property of ReqToCode: \textbf{traceability violations
are build failures.} The graduated lifecycle ensures this happens
predictably rather than abruptly.

\subsubsection{4.7 Change Detection --- Bidirectional
Visibility}\label{change-detection-bidirectional-visibility}

Consider a scenario where the title of \texttt{SWR-101} changes from
``Validate sensor range on input'' to ``Validate sensor range and unit
on input.'' The ALM tool records this change with a last-modified
timestamp of \texttt{2026-02-18T14:32:00Z}. The generated Traceable is
updated accordingly. The implementation code, however, still references
\texttt{SWR\_101} and compiles successfully --- the identifier has not
changed.

The system compares the requirement's last-modified timestamp from the
ALM tool (\texttt{2026-02-18}) against the last Git commit that touched
the referencing code (\texttt{2026-01-29}). The result: the requirement
was modified after its implementing code was last touched --- a
potential drift that has not been reviewed. This does not produce a
build error --- the trace is still structurally valid --- but it
produces a visibility signal that enables teams to assess whether the
implementation still satisfies the updated requirement.

The reverse case is equally detectable: if a developer commits a change
to the sensor validation implementation on \texttt{2026-02-20} but the
requirement's last-modified date in the ALM tool remains
\texttt{2026-01-15}, this mismatch becomes visible --- code changed, but
the requirement did not, suggesting either an undocumented behavioral
change or a missing requirement update.

\subsubsection{4.8 Branch-Scoped Traceability --- Parallel
Development}\label{branch-scoped-traceability-parallel-development}

Suppose the sensor validation component is part of a product baseline
maintained on the main branch. A new feature --- redundant sensor fusion
--- is being developed on a feature branch
\texttt{feature/sensor-fusion}. This feature introduces two additional
requirements:

{\def\LTcaptype{none} 
\begin{longtable}[]{@{}lll@{}}
\toprule\noalign{}
ID & Title & Status \\
\midrule\noalign{}
\endhead
\bottomrule\noalign{}
\endlastfoot
SWR-201 & Fuse readings from redundant sensors & Approved \\
SWR-202 & Detect sensor disagreement & Approved \\
\end{longtable}
}

The ReqToCode pipeline generates Traceables for all requirements ---
including SWR-201 and SWR-202 --- across all branches. Both the main
branch and the feature branch share the same complete set of Traceables.
The new requirements carry scope metadata that associates them with the
sensor fusion feature.

Developers working on the feature branch reference \texttt{SWR\_201} and
\texttt{SWR\_202} in their implementation and test code. These
references exist only on the feature branch. When the system reports
traceability back to the ALM tool for this branch, it reports only the
references that are new compared to main --- the feature's contribution
to the overall traceability state.

When the feature branch is merged into the main branch, no Traceable
conflicts arise --- the generated artifacts are identical on both
branches. Only the code references merge, reflecting the new
implementation and test coverage.

\subsubsection{4.9 Coverage Analysis --- On-Demand
Visibility}\label{coverage-analysis-on-demand-visibility}

At any point during development, a coverage analysis can be computed by
scanning the codebase for Traceable references. On the feature branch, a
branch-relative report shows only the delta --- the feature's
contribution:

{\def\LTcaptype{none} 
\begin{longtable}[]{@{}
  >{\raggedright\arraybackslash}p{(\linewidth - 6\tabcolsep) * \real{0.1341}}
  >{\raggedright\arraybackslash}p{(\linewidth - 6\tabcolsep) * \real{0.4268}}
  >{\raggedright\arraybackslash}p{(\linewidth - 6\tabcolsep) * \real{0.2927}}
  >{\raggedright\arraybackslash}p{(\linewidth - 6\tabcolsep) * \real{0.1463}}@{}}
\toprule\noalign{}
\begin{minipage}[b]{\linewidth}\raggedright
Traceable
\end{minipage} & \begin{minipage}[b]{\linewidth}\raggedright
Implementation References (delta)
\end{minipage} & \begin{minipage}[b]{\linewidth}\raggedright
Test References (delta)
\end{minipage} & \begin{minipage}[b]{\linewidth}\raggedright
Status
\end{minipage} \\
\midrule\noalign{}
\endhead
\bottomrule\noalign{}
\endlastfoot
SWR\_201 & 1 & 2 & Active \\
SWR\_202 & 1 & 1 & Active \\
\end{longtable}
}

On the main branch, the full baseline report covers all requirements:

{\def\LTcaptype{none} 
\begin{longtable}[]{@{}llll@{}}
\toprule\noalign{}
Traceable & Implementation References & Test References & Status \\
\midrule\noalign{}
\endhead
\bottomrule\noalign{}
\endlastfoot
SWR\_101 & 2 & 3 & Active \\
SWR\_102 & 1 & 1 & Deprecated \\
SWR\_103 & 0 & 0 & Active \\
\end{longtable}
}

This report reveals that \texttt{SWR\_103} has no implementation or test
references --- a gap that is immediately visible. It also shows that
\texttt{SWR\_102} is deprecated but still referenced, indicating cleanup
work in progress. After the feature branch is merged, \texttt{SWR\_201}
and \texttt{SWR\_202} appear in the main branch report as well. This
information is available on demand, at any commit, on any branch,
without manual data collection.

\begin{center}\rule{0.5\linewidth}{0.5pt}\end{center}

\subsection{5. Discussion}\label{discussion}

\subsubsection{5.1 Properties}\label{properties}

The ReqToCode approach provides the following properties when applied
consistently:

\textbf{Structural traceability.} Traces are code dependencies, not
metadata. They participate in the build process and are subject to the
same verification as any other code reference.

\textbf{Graduated lifecycle response.} Requirement changes propagate
through a deprecation-to-removal lifecycle, giving teams warning before
traces break. This makes the approach practical in environments where
requirement and code changes are not always synchronized.

\textbf{Immediate failure on removal.} When a Traceable is ultimately
removed, all dependent code produces compilation errors. There is no
state in which a requirement can be fully removed while its references
silently persist.

\textbf{Bidirectional change visibility.} Changes on either side of the
trace --- requirement or implementation --- are detectable through
revision comparison, enabling continuous alignment monitoring.

\textbf{Developer-native workflow.} Developers interact with Traceables
using their existing tools: IDE autocomplete, compiler diagnostics,
deprecation warnings, standard imports. No specialized traceability tool
is required during day-to-day development.

\textbf{Audit-readiness.} The trace structure is embedded in the version
control history. For any revision, the system can reconstruct which
Traceables existed, their lifecycle state, which were referenced in
code, and which were covered by tests. This provides a continuous,
Git-native audit trail.

\textbf{Branch-scoped traceability.} Different branches can carry
different sets of Traceables, reflecting distinct requirement scopes for
features, releases, or product variants. This enables parallel
development with isolated traceability state and meaningful merge
semantics.

\textbf{On-demand coverage analysis.} Requirement coverage --- both
implementation and test --- is computable at any revision on any branch
through static analysis of Traceable references. This eliminates manual
audit preparation and enables continuous coverage monitoring in CI/CD
pipelines.

\subsubsection{5.2 Limitations}\label{limitations}

\textbf{Trace granularity.} ReqToCode establishes that a requirement is
referenced in code but does not verify that the implementation is
correct or complete. A developer could reference \texttt{SWR\_101} in
code that does not actually implement range validation. The approach
guarantees structural presence, not semantic correctness.

\textbf{Adoption dependency.} The approach relies on developers
consistently referencing the generated Traceables. If a developer
implements a requirement without adding a trace reference, the system
cannot detect the omission. This risk can be mitigated through process
measures --- for example, by referencing requirements in implementation
tickets and tying the definition of done to the presence of Traceable
references in code. Such measures shift enforcement from the compiler to
the development workflow, which is less rigorous but practical in teams
that already follow structured processes. This limitation is shared with
all traceability approaches that do not employ full formal verification.

\textbf{Generation latency.} Since the pipeline synchronizes
periodically rather than in real time, there is a window between a
requirement change in the ALM tool and the corresponding update in the
generated Traceables. During this window, the codebase reflects the
previous state. The window is bounded and configurable, and in practice
has proven operationally preferable to real-time synchronization due to
reduced complexity.

\textbf{Language support.} The approach requires the target language to
support named constants, enumerations, or equivalent constructs that the
compiler can verify. Languages without compile-time verification of
named references (e.g., dynamically typed languages without build-time
checks) may require additional tooling to achieve equivalent guarantees.
Similarly, the deprecation lifecycle requires language-level deprecation
support; where this is absent, the system falls back to direct removal.

\textbf{ALM lifecycle exposure.} The graduated deprecation mechanism
depends on the ALM tool exposing requirement lifecycle transitions. Not
all tools provide this information through their APIs. Where lifecycle
data is unavailable, the system can only distinguish between
``requirement exists'' and ``requirement does not exist,'' reducing the
lifecycle to a binary active/removed model.

\subsubsection{5.3 Relationship to Recovery-Based
Approaches}\label{relationship-to-recovery-based-approaches}

ReqToCode and LLM-based traceability recovery {[}2, 3, 4{]} address the
same problem from opposite directions. Recovery approaches accept that
traces will degrade and invest in reconstructing them. ReqToCode invests
in preventing degradation structurally.

The two are complementary rather than competing. In a system that adopts
ReqToCode, LLM-based recovery could serve as a verification layer ---
identifying code that \emph{should} reference a Traceable but does not,
thereby addressing the adoption dependency limitation (Section 5.2).
Conversely, ReqToCode provides the structural anchors that recovery
approaches currently lack: rather than matching free-text requirement
descriptions to code through semantic similarity, a recovery tool
operating in a ReqToCode-enabled project could verify whether specific
Traceables are referenced where expected.

In addition to detecting missing references, LLM-based techniques could
also be applied to assess the semantic plausibility of existing trace
links. In a ReqToCode-enabled project, the presence of a Traceable
reference already establishes the structural relationship between
requirement and code. This bounded scope enables LLM-based analysis to
focus on a narrower question: whether the referencing code plausibly
implements the requirement's intent. Rather than attempting to discover
links in an unconstrained codebase, such techniques operate on
explicitly declared trace relationships, reducing the problem from link
discovery to plausibility assessment. Their role would therefore not be
to replace structural traceability, but to highlight ``suspect'' traces
for human review --- cases where a requirement is referenced
structurally but its semantic realization appears questionable.

\subsubsection{5.4 Implications for AI-Assisted
Development}\label{implications-for-ai-assisted-development}

The increasing use of AI code generation tools introduces a new
dimension to the traceability challenge. AI-generated code is typically
not linked to any requirement unless the developer explicitly
establishes the connection. As the volume of generated code grows, the
gap between what exists in the codebase and what is traceable widens.

Wang et al.~{[}5{]} propose embedding traceability as a first-class
objective in LLM code generation pipelines. ReqToCode provides the
structural foundation that such approaches require: the Traceables. If
an LLM generates code within a ReqToCode-enabled project, it can
reference the same Traceables as a human developer. The compile-time
verification then applies equally to AI-generated and handwritten code.
Code that does not reference a valid Traceable fails to build regardless
of its origin.

Beyond passive verification, Traceables can serve as \textbf{structured
context for LLM prompts}. Rather than instructing an LLM to ``implement
requirement SWR-101,'' a ReqToCode-enabled workflow can provide the
Traceable itself --- including its typed identifier, metadata, and
RequirementSet membership --- as part of the generation prompt. The LLM
receives not just a description but a concrete code element it must
reference in its output. This ties the generation act directly to the
trace structure, making the resulting code verifiable by the same
compile-time mechanisms that govern handwritten code.

This positions ReqToCode not only as a solution for current manual
traceability challenges but as an architectural foundation for
maintaining visibility in an era of increasing automation.

\subsubsection{5.5 Relationship to Existing
Standards}\label{relationship-to-existing-standards}

ReqToCode does not replace or implement any specific safety standard.
However, it supports the traceability objectives common to ISO 26262,
IEC 62304, DO-178C, and ASPICE by providing a mechanism that:

\begin{itemize}
\tightlist
\item
  Links requirements to implementation and test artifacts structurally
\item
  Detects broken or stale links automatically
\item
  Provides revision-level evidence of traceability state
\item
  Operates within the version control and build infrastructure that
  development teams already use
\end{itemize}

A concrete implementation of ReqToCode intended for use in
safety-critical development would itself require tool qualification
according to the applicable standard.

\subsubsection{5.6 Self-Application}\label{self-application}

The ReqToCode approach is not purely theoretical. A working
implementation --- \textbf{Ariadne} --- exists and is applied to its own
development process: Ariadne uses ReqToCode to trace its own
requirements. Ariadne currently generates language artifacts for Java,
C, and C++, and connects to Jira and Codebeamer as requirement sources,
as well as structured Markdown files for teams adopting a
requirements-as-code workflow. Generated Traceables,
RequirementSet-specific annotations, the deprecation lifecycle, and
coverage analysis are all exercised in the development of the system
itself.

This self-application serves two purposes. First, it validates the
approach under real conditions: requirement changes, branch-based
development, and continuous integration are not simulated but
encountered daily. Second, it produces the documentation and
traceability evidence that the tool itself would require for tool
qualification in a regulated context --- the system traces its own
compliance.

\begin{center}\rule{0.5\linewidth}{0.5pt}\end{center}

\subsection{6. Conclusion}\label{conclusion}

Requirements traceability in regulated software development has long
been treated as a documentation concern --- maintained externally,
verified periodically, and trusted to remain consistent with a system
that evolves continuously. This trust is routinely violated, as
empirical evidence confirms {[}1{]}.

Recent advances in LLM-based traceability recovery {[}2, 3{]} achieve
impressive accuracy in reconstructing broken traces, yet they accept
trace degradation as inevitable. ReqToCode proposes a different model:
traceability as a structural property of the codebase, enforced at
compile time and maintained continuously.

By generating language-native Traceables from authoritative requirement
sources and embedding them in the development workflow, the approach
creates hard, compile-time verifiable links between requirements and
their realization. The graduated lifecycle --- from active through
deprecation to removal --- ensures that requirement changes produce
proportional, actionable signals rather than abrupt disruption. Broken
traces become build failures. Divergence between intent and
implementation becomes visible continuously rather than during periodic
audits.

The approach is non-invasive, complementary to existing tooling, and
compatible with modern Git-based development workflows. It provides a
structural foundation that both human developers and AI code generation
tools can reference equally, addressing the emerging challenge of
maintaining traceability in systems where code is increasingly generated
rather than written.

As AI-assisted development accelerates the pace and volume of code
production, the need for structural traceability mechanisms will only
grow. ReqToCode provides a foundation for maintaining visibility and
accountability in systems where the connection between intent and
implementation must not only exist but be provable.

\begin{center}\rule{0.5\linewidth}{0.5pt}\end{center}

\subsection{Appendix A: Concrete Language Realization
(Java)}\label{appendix-a-concrete-language-realization-java}

To make the Traceable concept tangible, this appendix shows how the
pseudocode from Section 4 maps to a concrete language. In Java, a
RequirementSet is realized as an enumeration type, with each Traceable
as an enum constant carrying its metadata.

Traceable names are composed of the requirement identifier followed by
the requirement title, both normalized to valid constant names. This is
a deliberate design choice: the identifier prefix ensures uniqueness and
provides a direct reference to the ALM tool, while the title suffix
makes the Traceable discoverable through IDE autocomplete. A developer
typing \texttt{SWR\_101} finds the Traceable by ID; a developer typing
\texttt{VALIDATE\_SENSOR} finds it by intent. Both paths lead to the
same compile-time verifiable element.

\begin{Shaded}
\begin{Highlighting}[]
\KeywordTok{public} \KeywordTok{enum}\NormalTok{ SensorValidation\_SWR }\OperatorTok{\{}

  \FunctionTok{SWR\_101\_VALIDATE\_SENSOR\_RANGE\_ON\_INPUT}\OperatorTok{(}\StringTok{"SWR{-}101"}\OperatorTok{,}\NormalTok{ Status}\OperatorTok{.}\FunctionTok{APPROVED}\OperatorTok{),}
  \FunctionTok{SWR\_102\_REJECT\_STALE\_SENSOR\_READINGS}\OperatorTok{(}\StringTok{"SWR{-}102"}\OperatorTok{,}\NormalTok{ Status}\OperatorTok{.}\FunctionTok{APPROVED}\OperatorTok{),}
  \FunctionTok{SWR\_103\_LOG\_VALIDATION\_FAILURES}\OperatorTok{(}\StringTok{"SWR{-}103"}\OperatorTok{,}\NormalTok{ Status}\OperatorTok{.}\FunctionTok{DRAFT}\OperatorTok{);}

  \KeywordTok{private} \DataTypeTok{final} \BuiltInTok{String}\NormalTok{ requirementId}\OperatorTok{;}
  \KeywordTok{private} \DataTypeTok{final}\NormalTok{ Status status}\OperatorTok{;}

  \FunctionTok{SensorValidation\_SWR}\OperatorTok{(}\BuiltInTok{String}\NormalTok{ requirementId}\OperatorTok{,}\NormalTok{ Status status}\OperatorTok{)} \OperatorTok{\{}
    \KeywordTok{this}\OperatorTok{.}\FunctionTok{requirementId} \OperatorTok{=}\NormalTok{ requirementId}\OperatorTok{;}
    \KeywordTok{this}\OperatorTok{.}\FunctionTok{status} \OperatorTok{=}\NormalTok{ status}\OperatorTok{;}
  \OperatorTok{\}}
\OperatorTok{\}}
\end{Highlighting}
\end{Shaded}

When \texttt{SWR-102} enters a deprecated state, the generated output
changes:

\begin{Shaded}
\begin{Highlighting}[]
    \AttributeTok{@Deprecated}
\FunctionTok{SWR\_102\_REJECT\_STALE\_SENSOR\_READINGS}\OperatorTok{(}\StringTok{"SWR{-}102"}\OperatorTok{,}\NormalTok{ Status}\OperatorTok{.}\FunctionTok{DEPRECATED}\OperatorTok{),}
\end{Highlighting}
\end{Shaded}

Any code referencing \texttt{SWR\_102\_REJECT\_STALE\_SENSOR\_READINGS}
now produces a compiler warning. IDEs highlight deprecated references
visually and show them with strikethrough formatting, making affected
code immediately identifiable across the project. When the Traceable is
eventually removed from the enum, all references produce compilation
errors.

Implementation code references Traceables through generated,
RequirementSet-specific annotations:

\begin{Shaded}
\begin{Highlighting}[]
\AttributeTok{@TracesSWR}\OperatorTok{(}\NormalTok{SWR\_101\_VALIDATE\_SENSOR\_RANGE\_ON\_INPUT}\OperatorTok{)}
\KeywordTok{public} \BuiltInTok{Result} \FunctionTok{validateSensorInput}\OperatorTok{(}\NormalTok{SensorReading reading}\OperatorTok{)} \OperatorTok{\{}
  \ControlFlowTok{if} \OperatorTok{(}\NormalTok{reading}\OperatorTok{.}\FunctionTok{value}\OperatorTok{()} \OperatorTok{\textless{}}\NormalTok{ RANGE\_MIN }\OperatorTok{||}\NormalTok{ reading}\OperatorTok{.}\FunctionTok{value}\OperatorTok{()} \OperatorTok{\textgreater{}}\NormalTok{ RANGE\_MAX}\OperatorTok{)} \OperatorTok{\{}
    \ControlFlowTok{return}\NormalTok{ INVALID}\OperatorTok{;}
  \OperatorTok{\}}
  \ControlFlowTok{return}\NormalTok{ VALID}\OperatorTok{;}
\OperatorTok{\}}
\end{Highlighting}
\end{Shaded}

Test code uses a corresponding verification annotation:

\begin{Shaded}
\begin{Highlighting}[]
\AttributeTok{@Test}
\AttributeTok{@VerifiesSWR}\OperatorTok{(}\NormalTok{SWR\_101\_VALIDATE\_SENSOR\_RANGE\_ON\_INPUT}\OperatorTok{)}
\DataTypeTok{void} \FunctionTok{rejectsOutOfRangeInput}\OperatorTok{()} \OperatorTok{\{}
  \DataTypeTok{var}\NormalTok{ reading }\OperatorTok{=} \FunctionTok{createReading}\OperatorTok{(}\NormalTok{RANGE\_MAX }\OperatorTok{+} \DecValTok{1}\OperatorTok{);}
  \FunctionTok{assertThat}\OperatorTok{(}\FunctionTok{validateSensorInput}\OperatorTok{(}\NormalTok{reading}\OperatorTok{)).}\FunctionTok{isEqualTo}\OperatorTok{(}\NormalTok{INVALID}\OperatorTok{);}
\OperatorTok{\}}
\end{Highlighting}
\end{Shaded}

The annotations are RequirementSet-specific --- \texttt{@TracesSWR}
rather than a generic \texttt{@Traces} --- because Java annotations can
only reference a single enum type in their value parameter. This is a
language-level constraint, not a design choice. It has a practical
benefit: each annotation name makes the RequirementSet explicit at the
reference site. The annotations themselves are autogenerated alongside
the enum, so the constraint adds no manual effort. Furthermore, the
annotations use a source-level retention policy, meaning they are
evaluated by the compiler but are not retained in the compiled bytecode.
This ensures that Traceables serve their purpose entirely at build time
--- enforcing trace integrity and enabling static analysis --- without
adding any footprint to the deployed artifact.

Multiple Traceables can be referenced in a single annotation when a
method or test relates to more than one requirement:

\begin{Shaded}
\begin{Highlighting}[]
\AttributeTok{@TracesSWR}\OperatorTok{(\{}
\NormalTok{    SWR\_101\_VALIDATE\_SENSOR\_RANGE\_ON\_INPUT}\OperatorTok{,}
\NormalTok{    SWR\_102\_REJECT\_STALE\_SENSOR\_READINGS}
\OperatorTok{\})}
\KeywordTok{public} \BuiltInTok{Result} \FunctionTok{validateAndCheckStaleness}\OperatorTok{(}\NormalTok{SensorReading reading}\OperatorTok{)} \OperatorTok{\{} \KeywordTok{...} \OperatorTok{\}}
\end{Highlighting}
\end{Shaded}

In all cases, the Traceable name carries both the requirement ID and its
intent directly in the code. A developer reading
\texttt{@TracesSWR(SWR\_101\_VALIDATE\_SENSOR\_RANGE\_ON\_INPUT)} knows
both \emph{which} requirement is being traced and \emph{what} it demands
--- without consulting the ALM tool. The same pattern applies to other
languages with compile-time verifiable named constants. In C --- a
primary target for automotive embedded systems --- Traceables can be
realized as preprocessor macros or enum constants within a generated
header file. Trace references in code use macros that expand to no-ops
or are removed entirely through conditional compilation
(\texttt{\#ifdef\ TRACEABILITY\_ENABLED}). This gives teams compile-time
verification during development while producing binaries with zero
traceability overhead for production deployment --- a critical property
in resource-constrained embedded environments. In C++
(\texttt{enum\ class} with \texttt{{[}{[}deprecated{]}{]}}), Rust
(\texttt{enum} with \texttt{\#{[}deprecated{]}}), and similar languages,
equivalent patterns apply.

\begin{center}\rule{0.5\linewidth}{0.5pt}\end{center}

\subsection{References}\label{references}

{[}1{]} M. Ruiz, J. Y. Hu, and F. Dalpiaz, ``Why don't we trace? A study
on the barriers to software traceability in practice,''
\emph{Requirements Engineering}, vol.~28, pp.~619--637, 2023.

{[}2{]} F. Niu, R. Pan, L. C. Briand, and H. Hu, ``TVR: Automotive
system requirement traceability validation and recovery through
retrieval-augmented generation,'' \emph{arXiv preprint
arXiv:2504.15427}, 2025.

{[}3{]} T. Hey, D. Fuchß, J. Keim, and A. Koziolek, ``Requirements
traceability link recovery via retrieval-augmented generation,'' in
\emph{Proc. REFSQ}, 2025.

{[}4{]} J. Hassine, ``An LLM-based approach to recover traceability
links between security requirements and goal models,'' in \emph{Proc.
28th International Conference on Evaluation and Assessment in Software
Engineering (EASE)}, 2024, pp.~643--651.

{[}5{]} F. Wang et al., ``Embedding traceability in large language model
code generation: Towards trustworthy AI-augmented software
engineering,'' in \emph{Proc. 33rd ACM International Conference on the
Foundations of Software Engineering (FSE Companion)}, 2025.

{[}6{]} O. C. Z. Gotel and C. W. Finkelstein, ``An analysis of the
requirements traceability problem,'' in \emph{Proc. First International
Conference on Requirements Engineering}, 1994, pp.~94--101.

{[}7{]} A. D. Rodriguez, K. R. Dearstyne, and J. Cleland-Huang,
``Prompts matter: Insights and strategies for prompt engineering in
automated software traceability,'' in \emph{Proc. IEEE 31st
International Requirements Engineering Conference Workshops}, 2023,
pp.~455--464.

{[}8{]} J. H. Hayes, A. Dekhtyar, and S. K. Sundaram, ``Advancing
candidate link generation for requirements tracing: the study of
methods,'' \emph{IEEE Transactions on Software Engineering}, vol.~32,
no. 1, pp.~4--19, 2006.

{[}9{]} R. F. Paige et al., ``Rigorous identification and encoding of
trace-links in model-driven engineering,'' \emph{Software and Systems
Modeling}, vol.~10, no. 4, pp.~469--487, 2011.

{[}10{]} J. Cleland-Huang, O. Gotel, J. H. Hayes, P. Mäder, and A.
Zisman, ``Software traceability: Trends and future directions,'' in
\emph{Proc. Future of Software Engineering (FOSE)}, 2014, pp.~55--69.

{[}11{]} P. Mäder and O. Gotel, ``Towards automated traceability
maintenance,'' \emph{Journal of Systems and Software}, vol.~85, no. 10,
pp.~2205--2227, 2012.

{[}12{]} P. Mäder and A. Egyed, ``Do developers benefit from
requirements traceability when evolving and maintaining a software
system?,'' \emph{Empirical Software Engineering}, vol.~20, no. 2,
pp.~413--441, 2015.

{[}13{]} J. Guo, J. Cheng, and J. Cleland-Huang, ``Semantically enhanced
software traceability using deep learning techniques,'' in \emph{Proc.
39th IEEE/ACM International Conference on Software Engineering (ICSE)},
2017, pp.~3--14.

{[}14{]} G. Antoniol, G. Canfora, G. Casazza, A. De Lucia, and E. Merlo,
``Recovering traceability links between code and documentation,''
\emph{IEEE Transactions on Software Engineering}, vol.~28, no. 10,
pp.~970--983, 2002.

\begin{center}\rule{0.5\linewidth}{0.5pt}\end{center}

\emph{Preprint. Not peer-reviewed. February 2026.}

\emph{Correspondence: thorsten@ariadne-thread.io}

\end{document}